# INTEGRATING LLMs FOR GRADING AND APPEAL RESOLUTION IN COMPUTER SCIENCE EDUCATION

**Ilhan Aytutuldu**
Department of Computer Engineering
Gebze Technical University
Kocaeli, Turkiye
iaytutuldu@gtu.edu.tr

**Ozge Yol**
SSTTEK ACADEMY
İstanbul, Turkiye
ozge.yol@ssttek.com

**Yusuf Sinan Akgul**
Department of Computer Engineering
Gebze Technical University
Kocaeli, Turkiye
akgul@gtu.edu.tr

April 18, 2025

## ABSTRACT

This study explores the integration of Large Language Models (LLMs) into the grading and appeal resolution process in computer science education. We introduce AI-PAT, an AI-powered assessment tool that leverages LLMs to evaluate computer science exams, generate feedback, and address student appeals. AI-PAT was used to assess over 850 exam submissions and handle 185 appeal cases. Our multi-model comparison (ChatGPT, Gemini) reveals strong correlations between model outputs, though significant variability persists depending on configuration and prompt design. Human graders, while internally consistent, showed notable inter-rater disagreement, further highlighting subjectivity in manual evaluation. The appeal process led to grade changes in 74% of cases, indicating the need for continued refinement of AI evaluation strategies. While students appreciated the speed and detail of AI feedback, survey responses revealed trust and fairness concerns. We conclude that AI-PAT offers scalable benefits for formative assessment and feedback, but must be accompanied by transparent grading rubrics, human oversight, and appeal mechanisms to ensure equitable outcomes.

## 1   Introduction

With their capabilities of understanding, analyzing, generating, and synthesizing human language, LLMs have paved the way towards a new era in education and assessment. Innovations in educational technology strive for automating the labor-intensive and time-consuming processes of generating and analyzing textual content [1, 2]. LLM models (e.g., Generative Pre-trained Transformer - GPT), advanced in Natural Language Processing (NLP) and trained extensively with vast text data, can generate human-like content, answer questions, make recommendations, represent knowledge, and perform various language-related tasks with remarkable accuracy [3, 4].

Generative AI models (e.g., GPT variants) demonstrated proficiency in automated grading across various disciplines [5], showcasing their potential to enhance efficiency, accuracy, and transparency. LLMs can interpret and evaluate student responses in a context-sensitive manner, making them well-suited for complex assessment tasks like essay grading, as they can assess not only the content but also features such as coherence, structure, argumentation, accuracy, and appropriate terminology [6]. These tools excel in structured assessments such as multiple-choice question generation, but also create spaces for ongoing learning and continuous development. By providing automated feedback and prompting students for revising their work, AI models can assess student knowledge and abilities while fostering continuous improvement [7, 8]. Considering that programming assessments, code generation, and problem-solving require a deeper understanding and nuanced feedback, AI tools hold significant promise for advancing computer science education (e.g., software engineering or programming) [9]. Research has suggested that AI models, especially the Chat-GPT, function well in delivering personalized and effective learning experience through customized feedback and explanations to students [10].

Studies have shown that AI models can grade as accurately as human raters, particularly in essay assessment. LLMs, like GPT variants, have demonstrated high accuracy in evaluating K-12 students' responses to open-ended short-answer questions across various domains (e.g., science and history), closely aligning with expert human raters [11][12], in their study on AI exam graders (GPT-3.5, GPT-4, and Gemini-pro), also found these models to be reliable and trustworthy. To enhance grading accuracy, they employed structured rubrics and carefully crafted prompts, which minimized ambiguity and reduced inconsistencies. Their results showed that GPT-4 was particularly dependable, leading to recommendations for integrating AI into exam grading processes, especially for its efficiency in handling time-consuming evaluations. However, they emphasize that AI grading should be supplemented with human oversight, particularly for complex or subjective responses.

While research supports AI's potential in automated grading, scholars caution against over-reliance. Although models like GPT 3.5 and GPT 4 have shown strong alignment with human grading, particularly when provided with well-defined rubrics [13, 14, 15], challenges remain. These models struggle with tasks requiring context-specific evaluations [16] or multi-step reasoning, such as debugging complex algorithms and understanding operational semantics [17]. Moreover, their probabilistic nature can lead to inconsistent responses for identical queries. Advanced models like ChatGPT (GPT-4o) from OpenAI [18] Gemini 2 Pro from Google DeepMind, and deepseek-chat try to address concerns such as bias and opaque decision-making, offering pathways to grading accuracy and reliability [19]. Despite these advancements, AI-based assessments require careful oversight to mitigate limitations and fully realize its potential in education.

## 2 Study Objectives

In this study, we introduce an AI-powered assessment tool (AI-PAT) as an evaluation framework for grading computer science exams while measuring its effectiveness in automated grading and providing interactive, personalized feedback to students on programming tasks, code analysis, and theoretical questions. Additionally, we explore students' perceptions of AI-PAT to identify the framework's strengths and areas for improvement and propose strategies for effectively integrating AI into grading to foster student-centered learning.

The contribution of this study is threefold. Responding to the call by [20] for LLMs-based innovations that support classroom tasks and authentic, replicable studies, we introduce the AI-PAT process first. We present the example exams and student engagements with the AI when appealing their grades. Second, we introduce an assessment tool, AI-PAT, which is designed for seamless automation of grading not only of computer science exams but also for other university exams. AI-PAT can significantly reduce the arduous labor of manual evaluation and providing individualized feedback, allowing educators to focus more on impactful aspects of teaching. Finally, we believe that we are at the forefront of the change in knowledge and learning construction due to the advancements in LLMs. Representing a shift towards a more student-centered approach, AI-PAT fosters students' active engagement in their own learning and encourages students to critically analyze their exam responses. By prompting deeper reflection and structured argumentation during their grade appeal process, AI-PAT encourages students to take ownership of their own learning. Preparing students for this knowledge and learning transportation is crucial to enhance their academic development, and we believe AI-PAT is an important step for that.

In this study, we aim to respond these research questions:

1. To what extent do AI models differ in grading computer science exams?

2. To what extent do AI models differ from humans in grading computer science exams?

3. How effective is AI-PAT in resolving student appeals?

4. What are students' perceptions of AI-PAT?

## 3 AI-PAT: Instructional Procedures

In our OOP (Object-Oriented Programming) course, we integrated the AI-PAT as an evaluative framework. Our framework integrates secure handling of handwritten submissions, structured transcription, AI-assisted grading, and systematic feedback mechanisms to enhance learning outcomes. we developed and assessed nine quizzes, a midterm examination, and a final assessment for around 90 students. We created a single quiz question comprising two unique components: one that assesses comprehension and the other that evaluates problem solving, necessitating students to exhibit their understanding and application of essential subjects. The questions were designed to illustrate the principle of separation of concerns, shown by the division of headers and implementations into distinct parts of code. For midterm and final examinations, we broadened the scope to six to eight questions, facilitating a more thorough evaluation of the students' knowledge and comprehension of the course material. This methodology enabled a thorough analysis, with quizzes providing ongoing formative feedback and midterm and final examinations offering a cumulative evaluation of overall progress. Fig. 1 presents a flow chart that illustrates the workflow to integrate AI-PAT into the evaluation process, detailing the steps from secure submission handling to AI-powered assessment and feedback sharing.

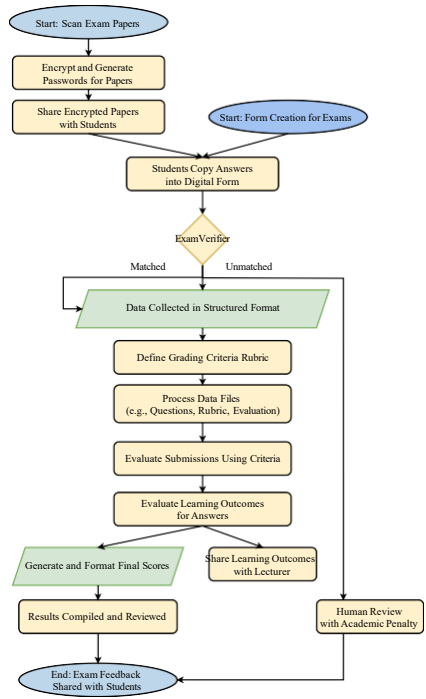

Figure 1: Workflow for integrating AI-PAT into the grading process for computer science exams. The process includes secure handling of handwritten submissions, structured transcription by students, AI-assisted grading using predefined rubrics, and detailed feedback sharing to enhance learning outcomes.

**ExamVerifier:** After each exam, when students transcribe their handwritten responses into digital format, there is a risk of intentional modifications or cheating. To ensure the authenticity of these digital submissions, our framework incorporates a script called ExamVerifier (Fig. 1). This script leverages Large Language Models (LLMs) to compare students' handwritten responses with their corresponding typed versions, detecting discrepancies and reinforcing academic integrity. If a mismatch is found, the submission is flagged for review, and in cases of intentional modifications (cheating), a penalty is applied.

We did not directly use handwritten OCR because even a single character can be crucial in evaluating our exams, especially in programming and technical subjects. Existing OCR technologies lack the necessary precision, often misinterpreting characters or symbols. Instead, our LLM-based approach focuses on identifying mismatches between handwritten and typed responses, providing a more reliable method for detecting potential cheating while ensuring fairness and accuracy in the grading process.

**Prompt Design:** After verifying student responses, we created an on-the-fly list of carefully designed prompts. This set includes the exam question, the student's answer to the question, clear grading guidelines, a structured output format to parse the model's response, a list of common mistakes and their possible penalties, and instructions for providing feedback while grading. [1]

In addition to the prepared prompt, we set the system role prompt as follows: 'You are a teaching assistant evaluating a student's answers on Object-Oriented Programming concepts in C++ or Java for correctness and completeness.' The exams were scored by LLMs using zero-shot learning with this role prompt.

Presented below is one of the midterm exam questions, designed to assess students' comprehension of copy constructors in C++. The question aims to evaluate their understanding of object-oriented programming principles, particularly regarding constructor correctness and function parameter passing.

---

[1] Some example scripts can be found at: https://github.com/iaytutu1/AI-powered-assessment-tool-AI-PAT.

**Question:** What is wrong with the following copy constructor declaration of the class `MyClass`? Provide an explanation for your answer.

```
MyClass(const MyClass o);
```

Table 1: Grading rubric for the midterm exam question on copy constructor correctness, generated by a large language model (LLM).

| Criteria | Total | Full Points | Partial Points | No Points |
|---|---|---|---|---|
| Identifying the Problem | 3 | Correctly identifies the issue, e.g., parameter passed by value instead of by reference, causing infinite recursion. | Recognizes a general issue (e.g., mentions inefficiency) but not the root cause. | Fails to identify the problem or provides an incorrect explanation. |
| Explaining Infinite Recursion | 3 | Clearly explains that passing `o` by value will call the copy constructor again, leading to infinite recursion and a stack overflow. | Provides an incomplete explanation, such as stating that recursion occurs but not explaining why or how. | Does not mention infinite recursion or provides an unrelated explanation. |
| Discussing Proper Copy Constructor Syntax | 2 | Explains that the parameter should be a reference (`const MyClass&`) to avoid recursion and improve efficiency. | Mentions the need for a reference but does not elaborate on why `const MyClass&` is preferred. | Does not mention proper syntax or provides an incorrect alternative. |

To ensure a structured evaluation, we employed a grading rubric that assesses students' ability to (i) identify the problem, (ii) explain the concept of infinite recursion, and (iii) discuss the correct syntax for a copy constructor. The grading rubric, generated by a large language model (LLM), is presented in Table 1.

The Python-based application integrated these prompts with LLMs via API. For each student response, the system formulated a custom prompt, invoking the API to evaluate the answer and assign a grade. The model simulated the role of a teaching assistant, analyzing correctness, clarity, and adherence to the specified criteria. The structured output, including detailed evaluations and assigned grades, were saved into an output spreadsheet for further review and dissemination.

**Challenge to LLMs:** To improve fairness, the study incorporates an appeal mechanism, allowing students to challenge AI-generated grades (Fig. 2). Students could raise objections or provide clarifications regarding their graded responses using Microsoft Forms. These appeals were collected and systematically reviewed. The objections, along with the specific context of the exam questions, were submitted to LLMs for further analysis and response generation. LLMs provided tailored replies to address each appeal, ensuring that students' concerns were resolved transparently and consistently. The responses were then shared with students for review.

## 4 Methods

This section describes the methodology used to respond to the research questions of the study.

### 4.1 Data collection and analysis

**Setting and Participants:** This study focused on assessing examinations for an Object-Oriented Programming (OOP) course in a second-year undergraduate program, encompassing long-answer questions and problem-solving tasks. A total of 102 students were registered for the course, consisting of 18 women and 84 men. Eleven undergraduate examinations, including quizzes, a midterm, and a final exam, were chosen from the Fall 2024 Computer Science program at Gebze Technical University. These assessments were chosen for their structured problem-solving nature, which aligns with AI-PAT's evaluation capabilities. The exams were conducted in an exam hall, written in English, and formatted to ensure consistency in evaluation.

**Data Sources:** In the AI-PAT grading process, we employed three principal data sources: (1) student examination papers in digital format, (2) student appeals accompanied with feedback on appeal resolutions, and (3) general student feedback regarding the entire system. These sources offered a thorough assessment of AI-PAT's efficacy, equity, and potential enhancements.

The procedure commenced with the gathering of handwritten examination submissions, which were subsequently scanned and encrypted via pyzipper for security purposes. Prior to distribution to students, each file was assigned a unique password to provide discreet access for reviewing their work and comprehending their performance. Students subsequently converted their handwritten responses into a structured digital version with Microsoft Forms [2], so ensuring uniformity for AI-assisted evaluation.

---

[2]Please see the form here: https://forms.office.com/r/XYdEpFpku7

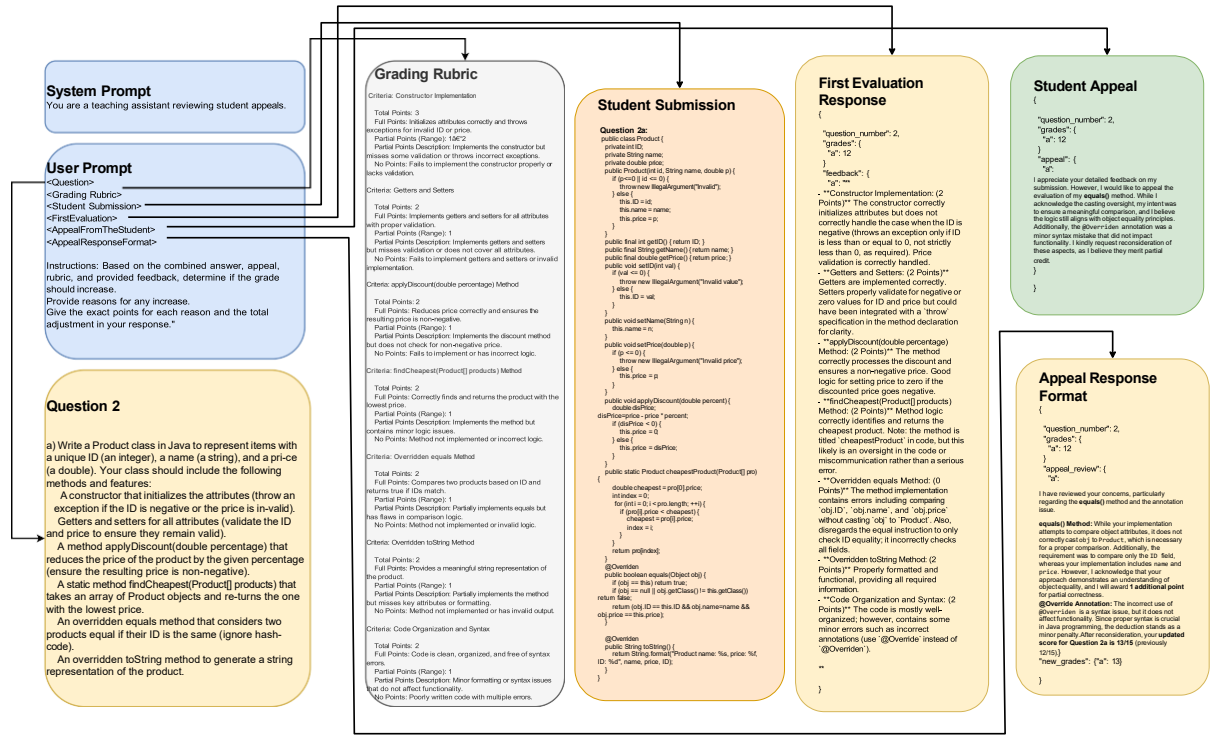

Figure 2: The diagram illustrates the sub-components of the appeal handling process, including the system prompt, the question, grading rubric, student submission, initial evaluation response, student appeal, and appeal review.

Upon the completion of the original grading procedure, we gathered student appeals to identify the existence of any grading discrepancies through the form [3]. Furthermore, we administered another survey [4] to assess the efficacy of the appeal resolution procedure. This poll collected student feedback regarding their appeal experiences, encompassing their views on the fairness of the settlement process and their happiness with the final conclusions.

At the end of the whole process, we gathered student feedback via a standardized questionnaire [5] to provide insights into their experiences with AI-PAT. The questionnaire sought to evaluate students' impressions of AI-generated grading, the clarity and utility of the feedback supplied, and their overall satisfaction with the system.

**Analysis of AI-PAT Grading:** We performed a thorough statistical analysis to evaluate the consistency and variability in grading among LLMs such as ChatGPT, Gemini, and DeepSeek. In the data analysis phase, we methodically examined the means and standard deviations of total paper scores among various LLMs. Due to the lack of contextual clarity in these metrics, we conducted a further assessment of inter-model relationships utilizing correlation coefficients—namely, the Pearson correlation coefficient (r) for linear relationships and Spearman's rank correlation coefficient for rank-order relationships between the grading systems.

Alongside assessing grading consistency among LLMs, we randomly selected a sample of 30 papers from a total of 73 final exam submissions for human evaluation. To evaluate intra-rater reliability (consistency of a single grader over time) and inter-rater reliability (consistency among several graders), two human evaluators independently assessed one particular question from the chosen papers. This question was chosen for its balanced evaluation of conceptual comprehension and coding expertise, necessitating students to express their thinking in an extended response format while also showcasing their programming abilities. One week later, the same evaluators reassessed the identical batch

---

of questions without access to their prior evaluations. This method enabled us to measure the consistency of individual assessors' grading over time and to compare the degree of concordance between the two evaluators in their evaluations.

**Evaluation of Appeal Resolutions and Student Feedback Analysis:** We performed an analysis of student appeals to evaluate the system's fairness and openness. This analysis investigated the prevalence of appeals, the categories of issues presented by students, and the outcomes of resolutions. We conducted a statistical comparison of the initial AI-generated grades and the changed scores post-appeal settlement to identify patterns in grading modifications. This comparison revealed potential biases, inconsistencies, and opportunities for further refinement of AI-PAT to improve grading reliability and student trust.

**Analysis of Student Perception on AI-PAT Grading:** We performed descriptive, comparative, and thematic analyses of questionnaire responses to assess students' perceptions of AI-PAT. Descriptive analysis yielded insights on overarching trends, including student satisfaction levels and preferences for AI-assisted grading. A comparative analysis investigated perceptual differences influenced by characteristics such as course repetition and assessment style, employing statistical tests to discern significant changes. The thematic analysis of open-ended replies identified significant concerns and recommendations, emphasizing issues such as equity, clarity of feedback, and precision in grading. Collectively, these evaluations provide a thorough knowledge of AI-PAT's influence on student experiences and identify areas for enhancement.

## 5 Results

### 5.1 Grading Performance Across LLMs: ChatGPT vs Gemini

**AI-PAT Grading Descriptive Statistics**

A summary of the grading distributions for ChatGPT and Gemini, including measures of central tendency and variability, is presented in Table 2.

Table 2: Descriptive Statistics for ChatGPT and Gemini in Final Exam Grades.

| Metric | Count | Mean | Std Dev | Min | 25% | Median | 75% | Max |
|--------|-------|------|---------|-----|-----|--------|-----|-----|
| gpt-4o | 73 | 36.11 | 17.22 | 8 | 21 | 36 | 49 | 77 |
| gemini-2.0-pro | 73 | 43.59 | 12.81 | 15 | 34 | 45 | 52 | 71 |

The statistics show that Gemini assigns higher average grades (43.59) compared to ChatGPT (36.11). However, ChatGPT has a higher standard deviation (17.22 vs. 12.81), meaning its grading is more spread out, while Gemini is more consistent. The distribution shows that Gemini scores tend to be higher, whereas ChatGPT has a wider spread of grades. Fig. 3 illustrates the grade distribution for ChatGPT and Gemini in final exam.

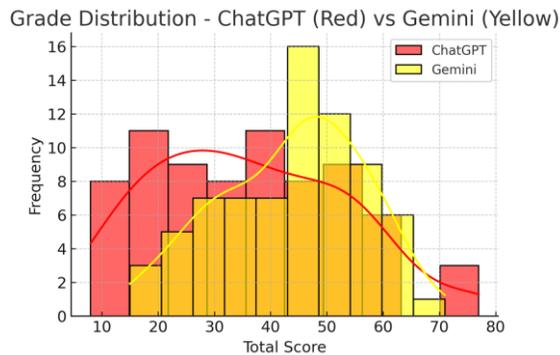

Figure 3: Distribution of Final Exam Grades for ChatGPT and Gemini.

**AI-PAT Grade-Wise Correlation Analysis**

To examine the grading consistency, we computed Pearson and Spearman correlation coefficients for each questions and the total score in final exam. The Pearson correlation values indicate a strong linear relationship between ChatGPT's and Gemini's grading. With a coefficient of 0.909, this suggests that students who score high in one system tend to score similarly in the other, demonstrating consistency between the two grading methods. The Spearman correlation is slightly higher in most cases, signifying that the ranking of students remains largely consistent between the two grading models. Table 3 presents the detailed correlation results.

Table 3: Correlation Coefficients Between ChatGPT and Gemini Scores for Each Grade Component and Total

| Grade Component (n = 73) | Pearson | Spearman |
|---|---|---|
| Question 1a | 0.722 | 0.681 |
| Question 1b | 0.729 | 0.753 |
| Grade 1c | 0.672 | 0.678 |
| Question 2a | 0.731 | 0.758 |
| Question 2b | 0.720 | 0.726 |
| Question 3a | 0.839 | 0.836 |
| Question 3b | 0.757 | 0.803 |
| Question 3c | 0.743 | 0.774 |
| Question 4 | 0.768 | 0.766 |
| TOTAL | 0.909 | 0.920 |

Table 4 presents the Pearson correlation coefficients (r) and p-values between ChatGPT and Gemini grades across different question components. Strong positive correlations (e.g., r=0.91 for total scores) indicate a high level of agreement between the two grading systems. Statistically significant correlations ($p<0.05$, marked with * ) suggest that as ChatGPT's grades increase, Gemini's grades tend to follow a similar trend, demonstrating consistency in evaluation.

Table 4: Pearson Correlation Coefficient r and (p-values). Star (*) indicates statistical significance at $p < 0.05$.

| Gemini Grades | ChatGPT Grades | | | | | | | | | |
|---|---|---|---|---|---|---|---|---|---|---|
| | Q1a | Q1b | Q1c | Q2a | Q2b | Q3a | Q3b | Q3c | Q4 | Total |
| Q1a | 0.72 (0.000*) | 0.53 (0.000) | 0.37 (0.001) | 0.24 (0.041) | 0.33 (0.004) | -0.06 (0.587) | 0.32 (0.005) | 0.18 (0.132) | 0.12 (0.315) | 0.38 (0.001) |
| Q1b | 0.49 (0.000) | 0.73 (0.000*) | 0.18 (0.130) | 0.25 (0.030) | 0.35 (0.002) | 0.09 (0.441) | 0.44 (0.000) | 0.14 (0.232) | 0.16 (0.172) | 0.43 (0.000) |
| Q1c | 0.41 (0.000) | 0.29 (0.014) | 0.67 (0.000*) | 0.34 (0.003) | 0.35 (0.002) | 0.04 (0.726) | 0.27 (0.019) | 0.13 (0.279) | 0.20 (0.092) | 0.37 (0.001) |
| Q2a | 0.35 (0.002) | 0.32 (0.007) | 0.40 (0.000) | 0.73 (0.000*) | 0.53 (0.000) | 0.41 (0.000) | 0.48 (0.000) | 0.39 (0.001) | 0.22 (0.067) | 0.68 (0.000) |
| Q2b | 0.25 (0.030) | 0.21 (0.077) | 0.38 (0.001) | 0.50 (0.000) | 0.71 (0.000*) | 0.41 (0.000) | 0.42 (0.000) | 0.26 (0.027) | 0.16 (0.000) | 0.63 (0.000) |
| Q3a | 0.08 (0.508) | 0.20 (0.087) | 0.11 (0.337) | 0.40 (0.000) | 0.39 (0.001) | 0.84 (0.000*) | 0.38 (0.001) | 0.51 (0.000) | 0.08 (0.481) | 0.59 (0.000) |
| Q3b | 0.27 (0.019) | 0.38 (0.001) | 0.17 (0.153) | 0.44 (0.000) | 0.49 (0.000) | 0.62 (0.000) | 0.75 (0.000*) | 0.60 (0.000) | 0.23 (0.047) | 0.76 (0.000) |
| Q3c | 0.11 (0.352) | 0.28 (0.018) | 0.06 (0.601) | 0.41 (0.000) | 0.34 (0.003) | 0.65 (0.000) | 0.55 (0.000) | 0.74 (0.000*) | 0.08 (0.505) | 0.63 (0.000) |
| Q4 | 0.23 (0.050) | 0.34 (0.003) | 0.29 (0.012) | 0.09 (0.425) | 0.22 (0.060) | 0.16 (0.171) | 0.31 (0.007) | 0.22 (0.056) | 0.77 (0.000*) | 0.42 (0.000) |
| Total | 0.43 (0.000) | 0.51 (0.000) | 0.40 (0.000) | 0.63 (0.000) | 0.67 (0.000) | 0.69 (0.000) | 0.74 (0.000) | 0.65 (0.000) | 0.36 (0.002) | 0.91 (0.000*) |

## 5.2 Grading Performance: Human Graders (HGs) vs LLMs

**Reliability and Consistency Analysis**

To assess the reliability and consistency of both human and AI grading, a coding question accounting for 15% of the final exam was independently evaluated by two HGs and two AI-based models (ChatGPT and Gemini) at two different time points (3 days apart). We intentionally selected graders with different levels of experience to examine how expertise impacts grading consistency and agreement. Human grader (HG) 1 has five semesters of experience in grading Object-Oriented Programming (OOP) assessments, while HG 2 has only one semester of experience in grading similar coursework (Table 5).

Table 5: Descriptive Statistics across HGs and Sessions

| | HG 1 - Session 1 | HG 1 - Session 2 | HG 2 - Session 1 | HG 2 - Session 2 |
|---|---|---|---|---|
| Mean | 6.80 | 7.13 | 9.00 | 8.77 |
| Std Dev | 3.16 | 3.12 | 3.53 | 3.54 |
| Min | 0.00 | 0.00 | 0.00 | 0.00 |
| Max | 14.00 | 15.00 | 15.00 | 15.00 |

Additionally, we evaluated the inter-rater reliability between the two AI models (ChatGPT and Gemini) to analyze whether they exhibit consistency in their grading trends and score assignments, providing insight into how different LLMs interpret the same rubric and student responses.

## Table 6: Reliability Matrix Colored by Pearson Correlation

| | ChatGPT-4o (Temp=1) Run1 | ChatGPT-4o (Temp=1) Run2 | ChatGPT-4o (Temp=0) Run1 | ChatGPT-4o (Temp=0) Run2 | ChatGPT-4 Turbo Run1 | ChatGPT-4 Turbo Run2 | ChatGPT-4 Turbo (Temp=0) Run1 | ChatGPT-4 Turbo (Temp=0) Run2 | Gemini Run1 | Gemini Run2 | HG 1 Session 1 | HG 1 Session 2 | HG 2 Session 1 | HG 2 Session 2 |
|---|---|---|---|---|---|---|---|---|---|---|---|---|---|---|
| ChatGPT-4o (Temp=1) Run1 | 1.000 | | | | | | | | | | | | | |
| ChatGPT-4o (Temp=1) Run2 | 0.648 | 1.000 | | | | | | | | | | | | |
| ChatGPT-4o (Temp=0) Run1 | 0.828 | 0.667 | 1.000 | | | | | | | | | | | |
| ChatGPT-4o (Temp=0) Run2 | 0.833 | 0.705 | 0.929 | 1.000 | | | | | | | | | | |
| ChatGPT-4 Turbo (Temp=1) Run1 | 0.646 | 0.568 | 0.634 | 0.659 | 1.000 | | | | | | | | | |
| ChatGPT-4 Turbo (Temp=1) Run2 | 0.542 | 0.563 | 0.664 | 0.630 | 0.824 | 1.000 | | | | | | | | |
| ChatGPT-4 Turbo (Temp=0) Run1 | 0.623 | 0.416 | 0.617 | 0.653 | 0.590 | 0.558 | 1.000 | | | | | | | |
| ChatGPT-4 Turbo (Temp=0) Run2 | 0.553 | 0.499 | 0.629 | 0.635 | 0.510 | 0.532 | 0.898 | 1.000 | | | | | | |
| Gemini Run1 | 0.399 | 0.563 | 0.443 | 0.427 | 0.683 | 0.606 | 0.424 | 0.393 | 1.000 | | | | | |
| Gemini Run2 | 0.436 | 0.513 | 0.486 | 0.460 | 0.738 | 0.647 | 0.438 | 0.400 | 0.976 | 1.000 | | | | |
| HG 1 Session 1 | 0.505 | 0.455 | 0.468 | 0.424 | 0.835 | 0.724 | 0.498 | 0.426 | 0.709 | 0.740 | 1.000 | | | |
| HG 1 Session 2 | 0.515 | 0.430 | 0.465 | 0.406 | 0.788 | 0.712 | 0.530 | 0.428 | 0.650 | 0.687 | 0.964 | 1.000 | | |
| HG 2 Session 1 | 0.254 | 0.343 | 0.169 | 0.214 | 0.557 | 0.450 | 0.344 | 0.326 | 0.573 | 0.580 | 0.495 | 0.429 | 1.000 | |
| HG 2 Session 2 | 0.272 | 0.385 | 0.211 | 0.253 | 0.573 | 0.470 | 0.337 | 0.358 | 0.580 | 0.587 | 0.508 | 0.431 | 0.987 | 1.000 |

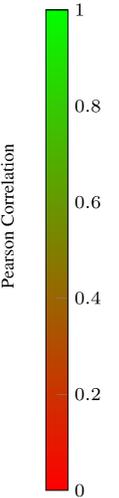

Pearson Correlation

The correlation matrix (Table 6) reveals clear patterns in grading reliability across different evaluators, particularly between HGs and LLM-based models. While some pairs show strong alignment, many others exhibit considerable divergence. Unsurprisingly, the highest consistency is observed between repeated runs of the same model configuration, especially under deterministic settings like zero temperature. For instance, ChatGPT-4o and Gemini both display high internal agreement between Run1 and Run2, suggesting that these models are relatively stable in their grading behavior when prompt, configuration, and randomness are controlled.

In contrast, HGs show more variability — not only between different human raters, but even within the same rater over two grading sessions. The correlation between HG 1's first and second attempts is high, but still notably less than the intra-model correlations. This implies that even trained HGs are subject to context drift, cognitive load, or subtle changes in interpretation over time. More critically, cross-rater reliability between HGs is weak, underscoring the subjective nature of manual assessment and the necessity of rubrics or training calibration sessions ( Table 6).

Among the LLMs, the different variants of ChatGPT and Gemini show moderate agreement across models, but not at levels that suggest full interchangeability. For example, changing the temperature parameter (e.g., from 0.7 to 0) within the same model leads to behavior shifts, and Pearson's correlations often drop into the range of 0.6 to 0.7. This highlights how model configuration can introduce inconsistency, even when using the same underlying architecture. Furthermore, models from different families (e.g. OpenAI ChatGPT vs. Google Gemini) align somewhat but never reach high-reliability thresholds, indicating that each LLM has a distinct grading 'style' or latent bias ( Table 6 ).

These findings suggest that neither human nor machine grading is inherently superior, but both are prone to inconsistency in different ways. For human evaluators, inconsistency may arise from cognitive fatigue, subjective interpretation, or ambiguity in rubrics — particularly among less experienced raters. Even experienced graders, however, can show variation across sessions due to contextual shifts or grading drift. On the other hand, large language models (LLMs) vary depending on prompt design, configuration settings like temperature, and inherent randomness in sampling. To mitigate these inconsistencies, grading systems should be designed with built-in redundancy and validation layers — such as ensemble evaluations or second-pass reviews, especially for borderline cases. Calibration sessions for human graders and fixed deterministic configurations for LLMs can help improve reliability. Ultimately, the variability observed in this matrix highlights the importance of establishing transparent, auditable, and fair grading pipelines that leverage the complementary strengths of both human expertise and machine consistency.

## 5.3 Appeal Resolutions

### Challenge to LLMs

Out of the 185 appeals submitted for approximately 850 evaluated papers, 137 out of 185 appeals (74%) resulted in grade changes, demonstrating that a significant proportion of initial evaluations were revised. However, 48 cases (26%) remained unchanged, indicating that some appeals did not provide sufficient grounds for modification.

Several students submitted appeals after the final exam, citing issues like zero credit despite partial correctness, syntax mix-ups (e.g., C++ and Java), and minor logic flaws. With the help of large language models (LLMs), these appeals were reviewed, and some grade adjustments were made to better reflect students' understanding and effort.

Table 7: Descriptive Analysis of Normalized Grades with Appeal Outcomes

|  | Original Grade | New Grade |
| --- | --- | --- |
| count | 185.00 | 185.00 |
| mean | 21.59 | 34.37 |
| std | 21.52 | 24.48 |
| min | 0.00 | 0.00 |
| max | 90.00 | 100.00 |
| Appeals with Change | 137.00 | 137.00 |
| Appeals with No Change | 48.00 | 48.00 |

Table 7 indicates the results of the students' challenge to LLMs. The statistical research indicates that quiz grades demonstrate the most variability after the appeal process (Table 8). This is probably attributable to the inherent characteristics of quizzes, which usually comprise a solitary question, rendering them more vulnerable to grading biases or misassessments. An insignificant error in assessment may result in a considerable disparity in the final score, given the absence of supplementary questions to equilibrate the evaluation. The elevated standard deviation noted in quiz grade modifications reinforces this idea, suggesting that alterations frequently lead to significant score fluctuations. Conversely, midterm and final examinations, which evaluate several facets of student performance, exhibit more moderate alterations following appeals.

Table 8: Statistical Analysis of Appealed Grades

| Exam Type | Original Mean (SD) | New Mean (SD) | Average Improvement |
|-----------|--------------------|--------------|--------------------|
| Quizzes | 18.73 (22.39) | 33.57 (26.77) | 14.84 |
| Midterm | 28.32 (17.36) | 34.02 (17.02) | 5.70 |
| Final | 31.24 (15.49) | 38.72 (15.20) | 7.48 |

The appeal process has been beneficial in enhancing grades for quizzes, midterms, and final examinations in appealing situations. The degree and range of modifications indicate possible discrepancies in the accuracy and complexity of original evaluations across these assessment categories. These findings underscore the demand for ongoing improvement of grading systems to augment fairness and consistency, hence diminishing the need for appeals in subsequent evaluations.

**Student Feedback on Appeal Resolutions**

Among the 185 appeals submitted for around 850 evaluated papers, feedback from 160 survey responses indicated varying degrees of satisfaction with the appeal process. Fig. 4 offers insights into students' perceptions of the appeal resolution procedure and the prospective application of AI-powered assessment tools. The fig. 4 (left) illustrates students' readiness to adopt AI for exam assessment, indicating differing degrees of confidence in automation. The fig. 4 (middle) illustrates their satisfaction with the appeal outcomes, reflecting whether they perceived their issues were treated equitably. The fig. 4 (right) analyzes the clarity of communication throughout the appeal process, elucidating the extent to which students comprehended the rationale behind the rulings. Collectively, these data provide an extensive perspective on students' experiences and anticipations regarding academic evaluation and conflict resolution.

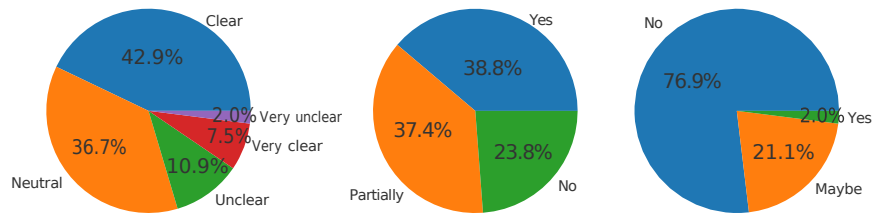

Figure 4: Survey results on student perceptions of AI-powered appeal resolutions. From left to right: Clarity of AI-PAT, Understanding AI-PAT, and Willingness to use AI-PAT for future assessments.

### 5.4 General Student Perception of AI-PAT

At the end of the semester, students were surveyed about their perceptions of AI-PAT, the AI-powered assessment tool used for grading in the course. The survey aimed to evaluate their attitudes toward AI-driven grading, the effectiveness of feedback, and their preference for future assessments. A total of 49 students participated in the survey.

The survey results (Fig. 5) from 49 students indicate a predominantly negative perception of the AI-Powered Assessment Tool (AI-PAT). While 12 students (24%) reported being somewhat satisfied, a larger portion expressed dissatisfaction, with 19 students (39%) somewhat dissatisfied and 18 students (37%) very dissatisfied. No students rated their experience as "very satisfied" or "neutral," suggesting a strong division between those who found some value in the system and those who did not.

Despite these concerns, 22 students (45%) agreed or strongly agreed that AI-PAT's feedback was helpful for their learning process. Students appreciated the speed of grading and the detailed feedback it provided. However, enthusiasm for the system did not translate into future preference, as only 6 students (12%) stated they would choose AI-PAT for future assessments, while 43 students (88%) opposed its continued use.

One of the key concerns raised was AI-PAT's inability to interpret responses fairly, particularly when handling syntax errors or minor mistakes. Some students felt that the system penalized small spelling errors or slight deviations from expected answers, leading to a perception of unfairness in grading. Trust in AI-PAT's impartiality was also extremely low, with zero students expressing confidence in its fairness. This skepticism was further reflected in how students compared AI-PAT to traditional grading methods—not a single respondent found AI-PAT better than human evaluation, indicating a widespread preference for instructor-led assessments.

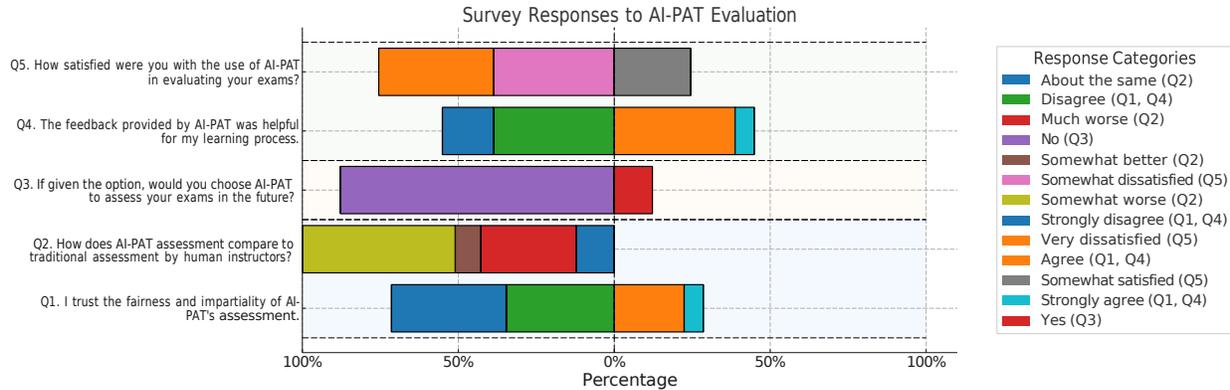

Figure 5: .

These results indicate that while students appreciate the speed and impartiality of AI grading, they still value human oversight, particularly when grading requires deeper contextual understanding. Further improvements in explainability and alignment with grading rubrics may increase trust in AI-PAT in future implementations.

# 6 Discussions

The integration of large language models (LLMs) into grading processes has raised significant discussions about their accuracy, reliability, and the impact on student learning outcomes. Automated grading systems powered by LLMs have demonstrated efficiency in handling large volumes of student responses while maintaining consistency in applying predefined grading rubrics [15, 21]. However, our results indicate that while LLMs maintain consistency in rubric application (Table 6), they struggle with subjective grading elements, mirroring findings from prior studies that highlight the difficulty of capturing human-like contextual judgment in open-ended assessments [22, 23].

One of the primary concerns in automated grading is the variability between different AI models. Our analysis found a high correlation between ChatGPT and Gemini scores (Table 4), but absolute score discrepancies remained, demonstrating model-dependent biases. Prior studies have similarly noted that LLMs tend to assign varying scores to identical responses, particularly in programming assessments where minor syntactical differences influence evaluations [6]. This suggests that further prompt refinement and score normalization techniques are necessary to mitigate these inconsistencies [24].

Intra-rater reliability, particularly among human graders, has traditionally been regarded as a benchmark for grading consistency [25]. Our study found that while experienced human graders demonstrated high self-consistency, ChatGPT exhibited more variability in repeated evaluations of the same answers, reinforcing concerns about the probabilistic nature of LLM outputs [26]. Similarly, [27] found that AI grading lacks stability over repeated iterations, emphasizing the importance of integrating periodic human calibration and structured rubrics to improve consistency.

The student appeal process has further highlighted critical limitations in AI-based grading. Our results indicate that a significant proportion of appeals led to grade revisions (Table 8), revealing potential grading inconsistencies in initial AI evaluations. This aligns with findings by [23, 22], which suggest that AI-based assessments require enhanced transparency to build student trust (Fig. 5). Our survey data (Fig. 4) supports this, as students expressed concerns about fairness and interpretability in AI grading, reinforcing the necessity for clearer AI-generated explanations.

Despite these challenges, our study found that AI-PAT not only expedited the grading process but also provided meaningful feedback that students found useful, as shown in Fig. 4, 5. While it has been demonstrated that the potential of LLMs to enhance feedback in programming education by improving code explanations and debugging guidance, our study highlights that AI-generated feedback must also prioritize grading consistency and transparency to ensure student trust and acceptance [28].

To enhance alignment and consistency, our research leverages approximately 850 exam papers and detailed grading criteria generated by ChatGPT, addressing the limitations observed in an earlier study conducted on a smaller sample of just 25 papers, where the need for refined criteria, clearer prompts, and calibration for both HGs and LLMs were highlighted [29].

[15] presents a compelling case for the role of well-designed prompts and structured rubrics in mitigating the limitations of AI-based assessment. Their study demonstrates that open-source models, when properly fine-tuned, can perform comparably to commercial alternatives, a finding that aligns with our observation that different LLMs exhibit similar trends in grading distribution (Fig. 3) [15]. However, skepticism toward AI grading remains high among students, as reflected in our survey data (Fig. 5), indicating that trust in automated systems is contingent on transparency and explainability [22].

## 7 Conclusions

This study demonstrates that LLMs can be effectively integrated into grading workflows, providing efficiency gains and structured feedback comparable to human graders. However, our results highlight key challenges, including grading variability, transparency concerns, and the necessity of human oversight. The student appeal process revealed inconsistencies that require further refinement, reinforcing the need for calibration techniques and enhanced prompt engineering. While open-source models show promise in achieving performance parity with commercial alternatives, their practical deployment depends on infrastructure capabilities. Future research should focus on improving AI-generated feedback clarity and developing hybrid grading models that combine the speed of LLMs with the contextual judgment of human evaluators. Addressing these concerns will be crucial for ensuring the broader acceptance of AI-powered grading in education.

## References

[1] Daniel Leiker, Ashley Ricker Gyllen, Ismail Eldesouky, and Mutlu Cukurova. Generative ai for learning: Investigating the potential of learning videos with synthetic virtual instructors. In *International conference on artificial intelligence in education*, pages 523–529. Springer, 2023.

[2] Sebastian Wollny, Jan Schneider, Daniele Di Mitri, Joshua Weidlich, Marc Rittberger, and Hendrik Drachsler. Are we there yet?-a systematic literature review on chatbots in education. *Frontiers in artificial intelligence*, 4:654924, 2021.

[3] Enkelejda Kasneci, Kathrin Seßler, Stefan Küchemann, Maria Bannert, Daryna Dementieva, Frank Fischer, Urs Gasser, Georg Groh, Stephan Günnemann, Eyke Hüllermeier, et al. Chatgpt for good? on opportunities and challenges of large language models for education. *Learning and individual differences*, 103:102274, 2023.

[4] Yan et al. Practical and ethical challenges of large language models in education: A systematic scoping review. *British Journal of Educational Technology*, 2023.

[5] Olaf Zawacki-Richter, Victoria I Marín, Melissa Bond, and Franziska Gouverneur. Systematic review of research on artificial intelligence applications in higher education–where are the educators? *International journal of educational technology in higher education*, 16(1):1–27, 2019.

[6] O Fagbohun, NP Iduwe, M Abdullahi, A Ifaturoti, and OM Nwanna. Beyond traditional assessment: Exploring the impact of large language models on grading practices. *Journal of Artifical Intelligence and Machine Learning & Data Science*, 2(1):1–8, 2024.

[7] Jürgen Rudolph, Shannon Tan, and Samson Tan. War of the chatbots: Bard, bing chat, chatgpt, ernie and beyond. the new ai gold rush and its impact on higher education. *Journal of Applied Learning and Teaching*, 6(1):364–389, 2023.

[8] Hong Ma and Tammy Slater. Using the developmental path of cause to bridge the gap between awe scores and writing teachers' evaluations. *Writing & Pedagogy*, 7(2-3):395–422, 2015.

[9] Roberto Rodriguez-Echeverría, Juan D Gutiérrez, José M Conejero, and Álvaro E Prieto. Analysis of chatgpt performance in computer engineering exams. *IEEE Revista Iberoamericana de Tecnologias del Aprendizaje*, 2024.

[10] Junaid Qadir. Engineering education in the era of chatgpt: Promise and pitfalls of generative ai for education. In *2023 IEEE global engineering education conference (EDUCON)*, pages 1–9. IEEE, 2023.

[11] Owen Henkel, Libby Hills, Bill Roberts, and Joshua McGrane. Can llms grade open response reading comprehension questions? an empirical study using the roars dataset. *International journal of artificial intelligence in education*, pages 1–26, 2024.

[12] Xinrui Song, Jiajin Zhang, Pingkun Yan, Juergen Hahn, Uwe Kruger, Hisham Mohamed, and Ge Wang. Integrating ai in college education: Positive yet mixed experiences with chatgpt. *Meta-Radiology*, 2(4):100113, 2024.

[13] Golchin et al. Grading massive open online courses using large language models. 2024.

[14] Mark J Gierl, Syed Latifi, Hollis Lai, André-Philippe Boulais, and André De Champlain. Automated essay scoring and the future of educational assessment in medical education. *Medical education*, 48(10):950–962, 2014.

[15] Policar et al. Automated assignment grading with large language models: Insights from a bioinformatics course. *Archive.org*, 2025.

[16] Jonas Flodén. Grading exams using large language models: A comparison between human and ai grading of exams in higher education using chatgpt. *British Educational Research Journal*, 2024.

[17] Chung Kwan Lo. What is the impact of chatgpt on education? a rapid review of the literature. *Education Sciences*, 13(4):410, 2023.

[18] Aaron Hurst, Adam Lerer, Adam P Goucher, Adam Perelman, Aditya Ramesh, Aidan Clark, AJ Ostrow, Akila Welihinda, Alan Hayes, Alec Radford, et al. Gpt-4o system card. *arXiv preprint arXiv:2410.21276*, 2024.

[19] Dinh et al. Sciex: Benchmarking large language models on scientific exams with human expert grading and automatic grading. 2024.

[20] Lixiang Yan, Lele Sha, Linxuan Zhao, Yuheng Li, Roberto Martinez-Maldonado, Guanliang Chen, Xinyu Li, Yueqiao Jin, and Dragan Gašević. Practical and ethical challenges of large language models in education: A systematic scoping review. *British Journal of Educational Technology*, 55(1):90–112, 2024.

[21] Flodén. Grading exams using large language models: A comparison between human and ai grading of exams in higher education using chatgpt. *British Educational Research Journal*, 2024.

[22] Fatih Yavuz, Özgür Çelik, and Gamze Yavaş Çelik. Utilizing large language models for efl essay grading: An examination of reliability and validity in rubric-based assessments. *British Journal of Educational Technology*, 56(1):150–166, 2025.

[23] Ryan Mok, Faraaz Akhtar, Louis Clare, Christine Li, Jun Ida, Lewis Ross, and Mario Campanelli. Using ai large language models for grading in education: A hands-on test for physics. *arXiv preprint arXiv:2411.13685*, 2024.

[24] Jiyeon Park and Sam Choo. Generative ai prompt engineering for educators: Practical strategies. *Journal of Special Education Technology*, page 01626434241298954, 2024.

[25] Marcus Messer, Neil CC Brown, Michael Kölling, and Miaojing Shi. How consistent are humans when grading programming assignments? *arXiv preprint arXiv:2409.12967*, 2024.

[26] Shahriar Golchin, Nikhil Garuda, Christopher Impey, and Matthew Wenger. Grading massive open online courses using large language models. *arXiv preprint arXiv:2406.11102*, 2024.

[27] Gustavo Pinto, Isadora Cardoso-Pereira, Danilo Monteiro, Danilo Lucena, Alberto Souza, and Kiev Gama. Large language models for education: Grading open-ended questions using chatgpt. In *Proceedings of the XXXVII brazilian symposium on software engineering*, pages 293–302, 2023.

[28] Mina Yousef, Kareem Mohamed, Walaa Medhat, Ensaf Hussein Mohamed, Ghada Khoriba, and Tamer Arafa. Begrading: large language models for enhanced feedback in programming education. *Neural Computing and Applications*, 37(2):1027–1040, 2025.

[29] Chokri Kooli and Nadia Yusuf. Transforming educational assessment: Insights into the use of chatgpt and large language models in grading. *International Journal of Human–Computer Interaction*, pages 1–12, 2024.